\documentclass{llncs} 
\usepackage{makeidx}
\usepackage{epsf}
\usepackage{latexsym}
\usepackage{amsmath, amssymb}
\usepackage{epsfig}
\begin{document}
\title{Forecasting Intermittent Demand by Hyperbolic-Exponential Smoothing}
\author{S. D. Prestwich${}^1$,
S. A. Tarim${}^2$,
R. Rossi${}^3$, 
and B. Hnich${}^4$\\ 
${}^1$Department of Computer Science, University College Cork, Ireland\\
${}^2$Department of Management, Hacettepe University, Ankara, Turkey\\
${}^3$University of Edinburgh Business School, Edinburgh, UK\\
${}^4$Computer Engineering Department, Izmir University of Economics, Turkey\\
{\tt s.prestwich@cs.ucc.ie},
{\tt armagan\_tarim@hacettepe.edu.tr},
{\tt roberto.rossi@wur.nl},
{\tt brahim.hnich@ieu.edu.tr}}
\institute{}
\maketitle
\begin{abstract}
Croston's method is generally viewed as superior to exponential
smoothing when demand is intermittent, but it has the drawbacks of
bias and an inability to deal with obsolescence, in which an item's
demand ceases altogether.  Several variants have been reported, some
of which are unbiased on certain types of demand, but only one recent
variant addresses the problem of obsolescence.  We describe a new
hybrid of Croston's method and Bayesian inference called
Hyperbolic-Exponential Smoothing, which is unbiased on
non-intermittent and stochastic intermittent demand, decays
hyperbolically when obsolescence occurs and performs well in
experiments.
\end{abstract}

\section{Introduction}


Inventory management is of great economic importance to industry, but
forecasting demand for spare parts is difficult because it is {\it
  intermittent\/}: in many time periods the demand is zero.  Various
methods have been proposed for forecasting, some simple and others
statistically sophisticated, but relatively little work has been done
on intermittent demand.


We now list the methods most relevant to this paper.  {\it Single
  exponential smoothing\/} (SES) generates estimates $\hat{y}_t$ of
the demand by exponentially weighting previous observations $y$ via
the formula
\[
\hat{y}_t = \alpha y_t + (1-\alpha)\hat{y}_{t-1}
\]
where $\alpha \in (0,1)$ is a {\it smoothing parameter\/}.  The
smaller the value of $\alpha$ the less weight is attached to the most
recent observations.  An up-to-date survey of exponential smoothing
algorithms is given in \cite{Gar}.  They perform remarkably well,
often beating more complex approaches \cite{FilEtc1}, but SES is known
to perform poorly (under some measures of accuracy) on intermittent
demand.

A well-known method for handling intermittency is {\it Croston's
  method\/} \cite{Cro} which applies SES to the demand sizes $y$ and
intervals $\tau$ independently.  Given smoothed demand $\hat{y}_t$ and
smoothed interval $\hat{\tau}_t$ at time $t$, the forecast is
\[
f_t = \frac{\hat{y}_t}{\hat{\tau}_t}
\]
Both $\hat{y}_t$ and $\hat{\tau}_t$ are updated at each time $t$ for
which $y_t \neq 0$.  According to \cite{Gar} it is hard to conclude
from the various studies that Croston's method is successful, because
the results depend on the data used and on how forecast errors are
measured.  But it is generally regarded as one of the best methods for
intermittent series \cite{GhoFri}, and versions of the method are used
in leading statistical forecasting software packages such as SAP and
Forecast Pro \cite{TeuEtc}.

To remove at least some of the known bias of Croston's method on
stochastic intermittent demand (in which demands occur randomly), a
correction factor is introduced by Syntetos \& Boylan \cite{SynBoy}:
\[
f_t = \left( 1 - \frac{\beta}{2} \right) \frac{\hat{y}_t}{\hat{\tau}_t}
\]
where $\beta$ is the smoothing factor used for inter-demand intervals,
which may be different to the $\alpha$ smoothing factor used for
demands.\footnote{In \cite{SynBoy} this factor is denoted by $\alpha$
  because it is used to smooth both $\hat{y}$ and $\hat{\tau}$.}
This works well for intermittent demand but is incorrect for
non-intermittent demand.  This problem is cured by Syntetos \cite{Syn}
who uses a forecast
\[
f_t = \left( 1-\frac{\beta}{2} \right)
\frac{\hat{y}_t}{\hat{\tau}_t-\frac{\beta}{2}}
\]
This reduces bias on non-intermittent demand, but slightly increases
forecast variance \cite{TeuSan}.

Another modified Croston method is given by Lev\'{e}n \& Segerstedt
\cite{LevSeg}, who claim that it also removes the bias in the original
method but in a simpler way: they apply SES to the ratio of demand
size and interval length each time a nonzero demand occurs.  That is,
they update the forecast using
\[
f_t = \alpha \left(\frac{y_t}{\tau_t}\right) + (1-\alpha) f_{t-1}
\]
However, this also turns out to be biased \cite{BoySyn}.

A more recent development is the TSB (Teunter-Syntetos-Babai)
algorithm \cite{TeuEtc}, which updates the demand probability instead
of the demand interval.  This allows it to solve the problem of {\it
  obsolescence\/} which was not previously dealt with in the
literature.  An item is considered obsolete if it has seen no demand
for a long time.  When many thousands of items are being handled
automatically, this may go unnoticed by Croston's method and its
variants.  One of the authors of this paper (Prestwich) has worked
with an inventory company who used Croston's method, but were forced
to resort to ad hoc rules such as: {\it if an item has seen no demand
  for 2 years then forecast 0.\/} TSB is designed to overcome this
problem.  Instead of a smoothed interval $\hat{\tau}$ it uses
exponential smoothing to estimate a probability $\hat{p}_t$ where
$p_t$ is 1 when demand occurs at time $t$ and 0 otherwise.  Different
smoothing factors $\alpha$ and $\beta$ are used for $\hat{y}_t$ and
$\hat{p}_t$ respectively.  $\hat{p}_t$ is updated every period, while
$\hat{y}_t$ is only updated when demand occurs.  The forecast is
\[
f_t=\hat{p}_t\hat{y}_t
\]



%

In this paper we shall use CR to denote the original method of
Croston, SBA the variant of Syntetos \& Boylan, SY that of Syntetos,
and TSB that of Teunter, Syntetos \& Babai.  We explore a new variant
of Croston's method that is unbiased and handles obsolescence.  Its
novelty is that during long periods of no demand its forecasts decay
hyperbolically instead of exponentially (as in TSB), a property that
derives from Bayesian inference.  The new method is described in
Section \ref{method} and evaluated in Section \ref{experiments}, and
conclusions are summarised in Section \ref{conclusion}.  This paper is
an extended version of \cite{PreEtc}.

\section{Hyperbolic-exponential smoothing} \label{method}



We take a Croston-style approach, separating demands into demand sizes
$y_t$ and the inter-demand interval $\tau_t$.  As in most Croston
methods, when non-zero demand occurs the estimated demand size
$\hat{y}_t$ and inter-demand period $\hat{\tau}_t$ are both
exponentially smoothed, using factors $\alpha$ and $\beta$
respectively.  The novelty of our method is what happens when there is
no demand.

Suppose that at time $t$, up to and including the last non-zero demand
we have smoothed demand size $\hat{y}_t$ and inter-demand period
$\hat{\tau}_t$, and that we have observed $\tau_t$ consecutive periods
without demand since the last non-zero demand.  What should be our
estimate of the probability that a demand will occur in the next
period?  A similar question was addressed by Laplace \cite{Lap}: given
that the sun has risen $N$ times in the past, what is the probability
that it will rise again tomorrow?  His solution was to add one to the
count of each event (the sun rising or not rising) to avoid zero
probabilities, and estimate the probability by counting the adjusted
frequencies.  So if we have observed $N$ sunrises and 0 non-sunrises,
in the absence of any other knowledge we would estimate the
probability of a non-sunrise tomorrow as $1/(N+2)$.  This is known as
the {\it rule of succession\/}.  But he noted that, given any
additional knowledge about sunrises, we should adjust this
probability.  These ideas are encapsulated in the modern {\it
  pseudocount\/} method which can be viewed as Bayesian inference with
a Beta prior distribution.  We base our discussion on a recent book
\cite{PooMac} (Chapter 7) that describes the technique we need in the
context of Bayesian classifiers.

For the two possibilities $y_t=0$ and $y_t \neq 0$ we add non-negative
pseudocounts $c_0$ and $c_1$ respectively to the actual counts $n_0$
and $n_1$ of observations.\footnote{These pseudocounts are often
  denoted $\alpha,\beta$ from the Beta distribution hyperparameters,
  but we already use these symbols for smoothing factors.}  As well as
addressing the problem of zero observations, pseudocounts allow us to
express the relative importance of prior knowledge and new data when
computing the posterior distribution.  By Bayes' rule the posterior
probability of a nonzero demand occurring is estimated by
\[
p(y_t \neq 0) = \frac{n_1 + c_1}{n_0 + c_0 + n_1 + c_1}
\]
(This is actually a conditional probability that depends on the recent
observations and prior probabilities, but we follow \cite{PooMac} and
write $p(y_t \neq 0)$ for simplicity.)  In our problem we have seen no
demand for $\tau_t$ periods so $n_1=0$ and $n_0= \tau_t$:
\[
p(y_t \neq 0) = \frac{c_1}{\tau_t + c_0 + c_1}
\]
We can eliminate one of the pseudocounts by noting that the prior
probability of a demand found by exponential smoothing is $1/
\hat{\tau}_t$, and that the pseudocounts must reflect this:
\[
\frac{c_1}{c_0 + c_1} = \frac{1}{\hat{\tau}_t}
\]
hence $c_0=c_1(\hat{\tau}_t-1)$ and
\[
p(y_t \neq 0) = \frac{c_1}{\tau_t + c_1 \hat{\tau}_t} = \frac{1}{\hat{\tau}_t+ \tau_t/c_1}
\]
As with TSB, to obtain a forecast we multiply this probability by the
smoothed demand size:
\[
f_t = \frac{\hat{y}_t}{\hat{\tau}_t+ \tau_t/c_1}
\]
We can also eliminate $c_1$ by choosing a value that gives an unbiased
forecaster on stochastic intermittent demand, as follows.  Consider
the demand sequence as a sequence of substrings, each starting with a
nonzero demand: for example the sequence $(5,0,0,1,0,0,0,3,0)$ has
substrings $(5,0,0)$, $(1,0,0,0)$ and $(3,0)$.  Within a substring
$\hat{y}_t$ and $\hat{\tau}_t$ remain constant so our forecaster
has expected forecast
\[
\mathbb{E}\left[ \frac{\hat{y}_t}{\hat{\tau}_t + \tau_t/c_1} \right]
= \mathbb{E}\left[ \frac{\hat{y}_t}{\hat{\tau}_t}
\left( \frac{1}{1 + \tau_t/ \hat{\tau}_tc_1} \right) \right]
\approx  \mathbb{E}\left[ \frac{\hat{y}_t}{\hat{\tau}_t}
\left( 1 - \frac{\tau_t} {\hat{\tau}_tc_1} \right) \right]
= \frac{\hat{y}_t}{\hat{\tau}_t} \left(1- \frac{1}{c_1}\right)
\]
The derivation used the linearity of expectation, the constancy of
$\hat{y}_t$ and $\hat{\tau}_t$, the fact that
$\mathbb{E}[\tau_t]=\hat{\tau}_t$, and the approximation $(1+
\delta)^{-1} \approx 1- \delta$ for small $\delta$ which can be found
by taking the first two terms of geometric series $(1+ \delta)^{-1}=
\sum_{i=0}^{\infty} \delta^i$.  Choosing $c_1=2/ \beta$, and therefore
$c_0=2(\hat{\tau}-1)/ \beta$, we obtain a forecast
\[
f_t = \frac{\hat{y}_t}{\hat{\tau}_t+ \beta\tau_t/2}
\]
with expected value
\[
\left(1- \frac{\beta}{2}\right) \frac{\hat{y}_t}{\hat{\tau}_t}
\]
which is identical to the actual SBA forecast.  So on any substring
our forecaster has the same expected forecast as SBA, given the same
values of $\hat{y}_t$ and $\hat{\tau}_t$.  Moreover, it updates
$\hat{y}_t$ and $\hat{\tau}_t$ in exactly the same way as SBA at the
start of each substring, therefore it has the same expected forecast
as SBA over the entire demand sequence.  Thus by \cite{SynBoy} it is
unbiased on stochastic intermittent demand.


A drawback with this forecaster is that, like SBA, it is biased on
non-intermittent demand.  This can be overcome by a slight adjustment
to the forecast:
\[
f_t = \frac{\hat{y}_t}{\hat{\tau}_t + \beta(\tau_t-1)/2}
\]
Following a similar derivation of the expected forecast:
\[
\begin{array}{rcl}
\mathbb{E}\left[ \frac{\hat{y}_t}{\hat{\tau}_t + \frac{\beta}{2}(\tau_t -1)} \right]
&=& \mathbb{E}\left[ \frac{\hat{y}_t}{\hat{\tau}_t}
\left( \frac{1}{1 + \frac{\beta}{2\hat{\tau}_t}(\tau_t -1)} \right) \right]\\
&\approx&  \mathbb{E}\left[ \frac{\hat{y}_t}{\hat{\tau}_t}
\left( 1 - \frac{\beta}{2\hat{\tau}_t}(\tau_t -1) \right) \right]\\
&=& \frac{\hat{y}_t}{\hat{\tau}_t} \left(1- \frac{\beta}{2}\right)
\left(1+ \frac{\beta}{2\hat{\tau}_t\left(1- \frac{\beta}{2}\right)} \right)\\
&\approx& \frac{\hat{y}_t}{\hat{\tau}_t} \left(1- \frac{\beta}{2}\right)
\left(1+ \frac{\beta}{2\hat{\tau}_t} \right)\\
&\approx& \left(1- \frac{\beta}{2}\right)
\frac{\hat{y}_t}{\hat{\tau}_t
\left(1 - \frac{\beta}{2\hat{\tau}_t} \right)}\\
&=& \left(1- \frac{\beta}{2}\right) \frac{\hat{y}_t}{\hat{\tau}_t - \frac{\beta}{2}}
\end{array}
\]
The final expression is exactly the forecast made by the SY method
throughout the substring.  But SY is unbiased on standard stochastic
intermittent demand and also on non-intermittent demand \cite{Syn}, so
(using the same arguments as above) our forecaster is too.  This is
the forecaster we shall use, and we call it Hyperbolic-Exponential
Smoothing (HES) because of its combination of exponential smoothing
with hyperbolic decay.

%

\begin{figure}
\begin{center}
\includegraphics[scale=0.9]{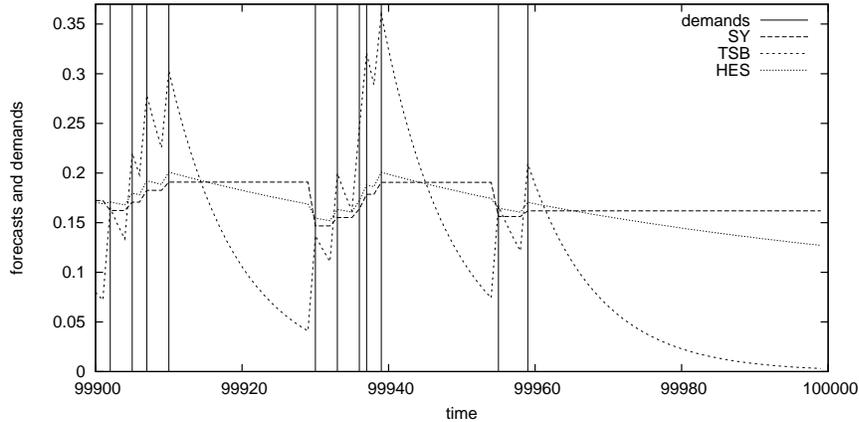}
\end{center}
\caption{Behaviour of SY, TSB and HES}
\label{pluseffect}
\end{figure}

An illustration of the different behaviour of SY, TSB and HES is shown
in Figure \ref{pluseffect}.  Demand is stochastic intermittent with
probability 0.25, all nonzero demands (shown as impulses) are 1, and
the forecasters use $\alpha=\beta=0.1$.  On stochastic intermittent
demand the HES forecasts are similar to those of SY, though there is
some decay between demands.  TSB also has greater variation, though
this difference could be reduced by using smaller smoothing
parameters.  When demand becomes zero, for example because the item
becomes obsolete, SY remains constant, TSB decays exponentially and
HES decays hyperbolically.


\section{Experiments} \label{experiments}

We now test HES empirically to evaluate its bias and forecasting
accuracy.  In all experiments, for CR, SBA and SY we let $\beta=
\alpha$ as is usual with those methods.

\subsection{Accuracy measures}

In any comparison of forecasting methods we must choose accuracy
measures.  \cite{GooHyn} lists 17 measures, noting that a
``bewildering array of accuracy measures have been used to evaluate
the performance of forecasting methods'', that no single method is
generally preferred, and that some are not well-defined on data with
intermittent demand.  We shall use measures that have been recently
recommended for intermittent demand.  To measure bias we use MASE
(Mean Absolute Scaled Error), recommended by \cite{HynKoe} and defined
as $\mbox{mean}(|q_t|)$ where $q_t$ is a scaled error defined by
\[
q_t = \frac{e_t}{\frac{1}{n-1} \sum_{i=2}^n |y_i-y_{i-1}|}
\]
$e_t$ is the error $y_t- \hat{y}_t$ and $t=1\ldots n$ are the time
periods of the samples used for forecasting, which we take to be the
$10^4$ samples used to initialise the smoothed estimates.  We take
these means over multiple runs.  MASE effectively evaluates a
forecasting method against the {\it naive\/} (or {\it random walk\/})
forecaster, which simply forecasts that the next demand will be
identical to the current demand.

As a measure of deviation we use the MAD/Mean Ratio (MMR), which has
been argued to be superior to several other methods used in
forecasting competitions \cite{KolSch} and is defined by
\[
\frac{\sum_{t=1}^n |e_t|}{\sum_{t=1}^n y_t}
\]
Again the summations are taken over multiple runs.

As another measure of deviation we also use the Relative Root Mean
Squared Error, defined as RMSE/RMSE${}_b$ where RMSE is measured on
the method being evaluated and RMSE${}_b$ on a baseline measure, both
taken over multiple runs.  When the baseline is random walk this is
Thiel's U2 statistic \cite{The}, and this is the baseline we use.  The
motivation behind using these two measures of deviation is that MMR is
based on absolute errors while U2 is based on root mean squared
errors; the latter penalises outliers more than the former so any
differences between them could be revealing.

For stationary demand experiments we shall also compare TSB and HES
using two relative measures.  Firstly we use the Relative Geometric
Root Mean Squared Error (RGRMSE) of HES with respect to TSB, analysed
by \cite{Fil1} and defined in our case as the geometric mean of
$|e_{\mbox{\tiny HES}}|/|e_{\mbox{\tiny TSB}}|$ taken over all periods
in multiple runs.  RGRMSE is also known as the Geometric Mean Relative
Absolute Error and was used by \cite{SynBoy}.  \cite{ArmCol}
recommends the use of such relative error-based measures, though they
have the drawback of potentially infinite variance because the
denominator can be arbitrarily small \cite{ChaHay,HynKoe}.  Secondly
we use the Percentage of times Better (PB), recommended by
\cite{KolSch} and defined as the percentage of times the absolute HES
error is less than the absolute TSB error.

\subsection{Logarithmic demand sizes} \label{biasstoch2}

We base our first experiments on those of Teunter {\it et al.\/}
\cite{TeuEtc} in which demands occur with some probability in each
period, hence inter-demand intervals are distributed geometrically,
and we use a logarithmic distribution for demand sizes.  Geometrically
distributed intervals are a discrete version of Poisson intervals, and
the combination of Poisson intervals and logarithmic demand sizes
yields a negative binomial distribution, for which there is
theoretical and empirical evidence: see for example the recent
discussion in \cite{SynEtc3}.

Teunter {\it et al.\/} compare several forecasters on demand that is
nonzero with probability $p_0$ where $p_0$ is either 0.2 or 0.5, and
whose size is logarithmically distributed.  The logarithmic
distribution is characterised by a parameter $\ell \in (0,1)$ and is
discrete with $\Pr[X=k]= - \ell^k/k \log(1- \ell)$ for $k \ge 1$.
They use two values: $\ell=0.001$ to simulate low demand and
$\ell=0.9$ to simulate lumpy demand.  They use $\alpha$ values 0.1,
0.2 and 0.3, and $\beta$ values 0.01, 0.02, 0.03, 0.04, 0.05, 0.1,
0.2, 0.3.  We add SY and HES to these experiments, but we drop SES as
they found it to have large errors.  They take mean results over 10
runs, each with 120 time periods, whereas we use 100 runs.  They
initialise each forecaster with ``correct'' values whereas we
initialise with arbitrary values $\hat{y}_0=\hat{\tau}_0=1$ then run
them for $10^4$ periods using demand probability $p_0$.  A final
difference is that instead of mean error and mean squared error we use
MASE and MMR/U2 respectively.

The results are shown in Tables \ref{sta1}--\ref{sta4}.  Because CR,
SBA and SY use only one smoothing factor $\alpha$ we do not show their
results for cases in which $\beta \neq \alpha$.  Comparing MASE
best-cases in each table, TSB and SY are least biased, closely
followed by HES, then CR and SBA.  Comparing MMR best-cases, SBA is
best, followed by TSB and HES, then CR and SY.  Comparing U2
best-cases HES is always at least as good as TSB, though there is
little difference.  Comparing MMR-worst cases, neither TSB nor HES
dominates the other though HES seems slightly better.  Again SBA gives
best results, CR and SY generally the worst.  Comparing U2-worst cases
HES beats TSB and seems to be more robust under different smoothing
factors.  SBA again gives best results, while CR and SY have variable
performance.

To examine the relative best-case performance of HES and TSB more
closely, Table \ref{logcomp} compares TSB and HES using RGRMSE and PB.
To make this comparison we must choose smoothing factors
$\alpha,\beta$ for both methods, and we do this in two different ways:
those giving the best MMR results and those giving the best U2
results.  The table also shows the best factors, denoted by
$\alpha_{\mbox{\tiny HES}},\beta_{\mbox{\tiny
    HES}},\alpha_{\mbox{\tiny TSB}},\beta_{\mbox{\tiny TSB}}$.  Using
the MMR-best factors HES performs rather less well than TSB on lumpy
demand, under both RGRMSE and PB.  But using U2-best factors there is
little difference between the methods.  The table also shows that to
optimise U2 we should use small factors for both TSB and HES, but to
optimise MMR the best factors depend on the form of the demand.

\subsection{Geometric demand sizes} \label{biasstoch3}

Another demand distribution that has recieved interest, and for which
there is also theoretical and empirical support, is the stuttering
Poisson distribution \cite{SynEtc3} in which demand intervals are
Poisson and demand sizes are geometrically distributed.  Again we use
a discrete version with geometrically distributed intervals.  The
geometric distribution is characterised by a probability we shall
denote by $g$, and is discrete with $\Pr[X=k]=(1-g)^{k-1}g$ for $k \ge
1$.  We use two values of $g$: 0.2 and 0.8 to simulate low and lumpy
demand respectively.  Otherwise the experiments are as in Section
\ref{biasstoch2}.

The results are shown in Tables \ref{stu1}--\ref{stu4}.  Though the
numbers are different, qualitatively the TSB and HES results are the
same as for logarithmic demand sizes.  However, CR, SBA and SY now give
similar results to each other, and SBA no longer has the best best-case
U2 result, though it still has the best worst-case U2 result.  Table
\ref{geocomp} compares TSB and HES using the RGRMSE and PB measures.
Qualitatively the results are the same as with logarithmic demand
sizes, except that HES is now worse than TSB in all 4 tables for MMR
best-cases --- though there is still little difference between the
U2-best cases.

\subsection{Decreasing demand} \label{decreasing}

The experiments so far use stationary demand, but Teunter {\it et
  al.\/} also consider nonstationary demand.  Again demand sizes
follow the logarithmic distribution, while the probability of a
nonzero demand decreases linearly from $p_0$ in the first period to 0
during the last period.  Demand sizes are again logarithmically
distributed.  As pointed out by Teunter {\it et al.\/}, none of the
forecasters use trending to model the decreasing demand so all are
positively biased.

The results are shown in Tables \ref{dec1}--\ref{dec4}.  TSB clearly
has the best best-case MASE, MMR and U2 results, while HES is
next-best, though SBA occasionally beats HES.  HES has the worst
worst-case MASE, MMR and U2 results, followed by TSB.  However, if we
only consider HES and TSB results for which $\alpha= \beta$ then CR,
SBA and SY have the worst worst-cases, with HES also poor on lumpy
demand.

The best $\beta$ for all methods is larger: as pointed out by Teunter
{\it et al.\/}, large smoothing factors are best at handling
non-stationary demand, while small factors are best when demand is
stationary.  The best $\alpha$ value is relatively unimportant here,
which makes sense as demand sizes are stationary.

\subsection{Sudden obsolescence} \label{sudden}

We repeat the experiments of Section \ref{decreasing}, but instead of
linearly decreasing the probability of demand we reduce it immediately
to 0 after half (60) of the periods, again following Teunter {\it et
  al.\/} Demand sizes are again logarithmically distributed.  The
results are shown in Tables \ref{obs1}--\ref{obs4}.  As found by
Teunter {\it et al.\/} the differences between TSB and CR, SBA and SY
are more pronounced because the latter are given no opportunity to
adjust to the change in demand pattern.  The results are qualitatively
similar to those for decreasing demand, but with greater differences.

\subsection{Bias at issue points only}

It is noted in \cite{TeuEtc} that although TSB is unbiased in the
above sense, it is biased if we only compare forecasts with expected
demand at issue points only (that is when demand occurs).  SES is
similarly biased, but not Croston methods such as SBA or SY.  The
cause is the decay in forecast size between demands, and HES will
clearly suffer from a similar bias.  We repeated the stationary demand
experiments with logarithmically and geometrically distributed demand
sizes, and measured the bias of TSB and HES based on issue points
only.  Both had greater bias than SY (as expected) but neither
dominated the other.

\subsection{Summary of empirical results}

On stationary demand SBA performs very well, followed by TSB and HES.
The relative performance of HES and TSB depends on how they are tuned.
If we tune them using MMR then TSB beats HES under both the RGRMSE and
PB measures, but the best smoothing factors are erratic; while if we
tune them using U2 there is no significant difference between them,
and the best smoothing factors are small.  We prefer to take the
U2-based results, not because they are better for our method but
because they are more consistent with other work: Teunter {\it et
  al.\/} found that small smoothing factors are best for stationary
demand, though admittedly this could simply be because they used
another (unscaled) mean squared error measure.  Intuitively this makes
sense, whereas using MMR we found no consistent results for the best
smoothing factors.  \cite{GhoFri} also recommend tuning by U2.  Thus
we recommend tuning forecasting methods using U2 rather than MMR, and
when doing this small smoothing factors are best for stationary
demand.  Under these conditions TSB and HES seem to be equally good
under two different measures, though they are slightly beaten by SBA.

HES is more robust than TSB under changes to the smoothing factors,
with better worst-case behaviour as measured by both MMR and U2.  We
believe that this is because HES's hyperbolic decay between demands is
slower than TSB's exponential decay, so large smoothing factors are
less harmful.  But on non-stationary demand, with intervals increasing
linearly or abruptly, TSB is best followed by HES, with CR, SBA and SY
giving worse performance if we consider the same range of smoothing
factor settings.  Here TSB's greater reactivity serves it well.  As
found by other researchers, and as is intuitively clear, large
smoothing factors are best at handling changes in demand pattern.
However, HES's robustness means that we can recommend smoothing
factors that behave reasonably well on both stationary and
non-stationary demand: $\alpha= \beta=0.1$.  The results in all cases
are not much worse than with optimally-tuned factors.

\section{Conclusion} \label{conclusion}

We presented a new forecasting method called Hyperbolic-Exponential
Smoothing (HES), which is based on an application of Bayesian
inference when no demand occurs.  We showed theoretically that HES is
approximately unbiased, and compared it empirically with Croston
variants CR, SBA, SY and TSB.  On stationary demand we found little
difference between TSB and HES, though HES was more robust under
changes to smoothing factors, and both performed well against other
Croston methods.  On non-stationary demand TSB performed best,
followed by HES.

Like TSB, HES has two smoothing factors $\alpha$ and $\beta$.  In
common with other methods, HES performs best on stationary demand with
small smoothing factors, and best on non-stationary demand with larger
factors.  Using smoothing factors $\alpha= \beta=0.1$ gave reasonable
results on a variety of demand patterns and we recommend these values.



\subsubsection*{Acknowledgment}

Thanks to Aris Syntetos for helpful advice, and to the anonymous
referees for useful criticism.  This work was partially funded by
Enterprise Ireland Innovation Voucher IV-2009-2092.

\bibliographystyle{plain}

\appendix

\section{Results}

\begin{table}
\begin{center}
\begin{tabular}{|rr|rrr|rrr|rrr|}
\hline
&& \multicolumn{3}{c|}{CR} & \multicolumn{3}{c|}{SBA} & \multicolumn{3}{c|}{SY}\\
$\alpha$ & $\beta$ & MASE & MMR & U2 & MASE & MMR & U2 & MASE & MMR & U2\\
\hline
0.10 & 0.10 & 0.015 & 1.219 & 0.717 & -0.019 & 1.194 & 0.717 & -0.001 & 1.207 & 0.717\\
0.20 & 0.20 & 0.032 & 1.249 & 0.730 & -0.037 & 1.197 & 0.726 & -0.000 & 1.225 & 0.728\\
0.30 & 0.30 & 0.051 & 1.283 & 0.745 & -0.056 & 1.200 & 0.735 & 0.003 & 1.247 & 0.741\\
\hline
\end{tabular}
\end{center}
\begin{center}
\begin{tabular}{|rr|rrr|rrr|}
\hline
&& \multicolumn{3}{c|}{TSB} & \multicolumn{3}{c|}{HES}\\
$\alpha$ & $\beta$ & MASE & MMR & U2 & MASE & MMR & U2\\
\hline
0.10 & 0.01 & -0.006 & 1.200 & 0.715 & -0.007 & 1.198 & 0.715\\
0.10 & 0.02 & -0.004 & 1.203 & 0.716 & -0.006 & 1.200 & 0.715\\
0.10 & 0.03 & -0.003 & 1.204 & 0.716 & -0.005 & 1.201 & 0.715\\
0.10 & 0.04 & -0.002 & 1.206 & 0.717 & -0.004 & 1.202 & 0.716\\
0.10 & 0.05 & -0.002 & 1.207 & 0.717 & -0.003 & 1.203 & 0.716\\
0.10 & 0.10 & -0.001 & 1.211 & 0.720 & -0.001 & 1.207 & 0.717\\
0.10 & 0.20 & -0.000 & 1.223 & 0.726 & 0.001 & 1.212 & 0.720\\
0.10 & 0.30 & -0.000 & 1.236 & 0.732 & 0.004 & 1.219 & 0.722\\
\hline
0.20 & 0.01 & -0.007 & 1.211 & 0.723 & -0.009 & 1.209 & 0.722\\
0.20 & 0.02 & -0.005 & 1.214 & 0.723 & -0.007 & 1.211 & 0.723\\
0.20 & 0.03 & -0.004 & 1.215 & 0.724 & -0.006 & 1.213 & 0.723\\
0.20 & 0.04 & -0.004 & 1.216 & 0.725 & -0.005 & 1.214 & 0.723\\
0.20 & 0.05 & -0.003 & 1.218 & 0.725 & -0.005 & 1.214 & 0.724\\
0.20 & 0.10 & -0.003 & 1.223 & 0.728 & -0.003 & 1.218 & 0.725\\
0.20 & 0.20 & -0.002 & 1.234 & 0.734 & -0.001 & 1.224 & 0.728\\
0.20 & 0.30 & -0.002 & 1.247 & 0.741 & 0.002 & 1.230 & 0.731\\
\hline
0.30 & 0.01 & -0.007 & 1.224 & 0.731 & -0.009 & 1.222 & 0.731\\
0.30 & 0.02 & -0.005 & 1.226 & 0.732 & -0.007 & 1.224 & 0.731\\
0.30 & 0.03 & -0.005 & 1.227 & 0.732 & -0.006 & 1.225 & 0.732\\
0.30 & 0.04 & -0.004 & 1.229 & 0.733 & -0.006 & 1.226 & 0.732\\
0.30 & 0.05 & -0.004 & 1.230 & 0.734 & -0.005 & 1.227 & 0.732\\
0.30 & 0.10 & -0.003 & 1.235 & 0.737 & -0.003 & 1.230 & 0.734\\
0.30 & 0.20 & -0.003 & 1.247 & 0.744 & -0.001 & 1.236 & 0.737\\
0.30 & 0.30 & -0.002 & 1.259 & 0.751 & 0.002 & 1.243 & 0.740\\
\hline
\end{tabular}
\end{center}
\caption{Logarithmic demand sizes with $\ell=0.9,p_0=0.5$}
\label{sta1}
\end{table}

\begin{table}
\begin{center}
\begin{tabular}{|rr|rrr|rrr|rrr|}
\hline
&& \multicolumn{3}{c|}{CR} & \multicolumn{3}{c|}{SBA} & \multicolumn{3}{c|}{SY}\\
$\alpha$ & $\beta$ & MASE & MMR & U2 & MASE & MMR & U2 & MASE & MMR & U2\\
\hline
0.10 & 0.10 & 0.027 & 1.651 & 0.707 & -0.000 & 1.617 & 0.707 & 0.005 & 1.624 & 0.707\\
0.20 & 0.20 & 0.053 & 1.695 & 0.714 & -0.004 & 1.621 & 0.711 & 0.008 & 1.637 & 0.712\\
0.30 & 0.30 & 0.082 & 1.746 & 0.723 & -0.008 & 1.626 & 0.717 & 0.013 & 1.654 & 0.719\\
\hline
\end{tabular}
\end{center}
\begin{center}
\begin{tabular}{|rr|rrr|rrr|}
\hline
&& \multicolumn{3}{c|}{TSB} & \multicolumn{3}{c|}{HES}\\
$\alpha$ & $\beta$ & MASE & MMR & U2 & MASE & MMR & U2\\
\hline
0.10 & 0.01 & 0.007 & 1.623 & 0.706 & 0.014 & 1.631 & 0.706\\
0.10 & 0.02 & 0.004 & 1.621 & 0.707 & 0.012 & 1.629 & 0.706\\
0.10 & 0.03 & 0.002 & 1.620 & 0.707 & 0.010 & 1.627 & 0.706\\
0.10 & 0.04 & 0.002 & 1.620 & 0.708 & 0.008 & 1.625 & 0.706\\
0.10 & 0.05 & 0.002 & 1.621 & 0.709 & 0.007 & 1.624 & 0.706\\
0.10 & 0.10 & 0.003 & 1.627 & 0.712 & 0.004 & 1.622 & 0.707\\
0.10 & 0.20 & 0.003 & 1.639 & 0.720 & 0.005 & 1.624 & 0.708\\
0.10 & 0.30 & 0.002 & 1.651 & 0.728 & 0.008 & 1.631 & 0.709\\
\hline
0.20 & 0.01 & 0.009 & 1.634 & 0.709 & 0.016 & 1.642 & 0.709\\
0.20 & 0.02 & 0.005 & 1.630 & 0.710 & 0.014 & 1.640 & 0.709\\
0.20 & 0.03 & 0.004 & 1.628 & 0.711 & 0.012 & 1.638 & 0.709\\
0.20 & 0.04 & 0.003 & 1.628 & 0.711 & 0.010 & 1.636 & 0.709\\
0.20 & 0.05 & 0.003 & 1.629 & 0.712 & 0.009 & 1.635 & 0.709\\
0.20 & 0.10 & 0.003 & 1.634 & 0.716 & 0.006 & 1.632 & 0.710\\
0.20 & 0.20 & 0.003 & 1.646 & 0.724 & 0.006 & 1.633 & 0.711\\
0.20 & 0.30 & 0.003 & 1.657 & 0.733 & 0.009 & 1.638 & 0.713\\
\hline
0.30 & 0.01 & 0.010 & 1.643 & 0.713 & 0.018 & 1.652 & 0.713\\
0.30 & 0.02 & 0.007 & 1.638 & 0.714 & 0.016 & 1.650 & 0.713\\
0.30 & 0.03 & 0.005 & 1.637 & 0.714 & 0.014 & 1.647 & 0.713\\
0.30 & 0.04 & 0.004 & 1.636 & 0.715 & 0.012 & 1.645 & 0.713\\
0.30 & 0.05 & 0.003 & 1.636 & 0.715 & 0.011 & 1.644 & 0.713\\
0.30 & 0.10 & 0.003 & 1.641 & 0.719 & 0.008 & 1.640 & 0.714\\
0.30 & 0.20 & 0.003 & 1.652 & 0.728 & 0.007 & 1.641 & 0.715\\
0.30 & 0.30 & 0.002 & 1.663 & 0.738 & 0.010 & 1.646 & 0.717\\
\hline
\end{tabular}
\end{center}
\caption{Logarithmic demand sizes with $\ell=0.9,p_0=0.2$}
\label{sta2}
\end{table}

\begin{table}
\begin{center}
\begin{tabular}{|rr|rrr|rrr|rrr|}
\hline
&& \multicolumn{3}{c|}{CR} & \multicolumn{3}{c|}{SBA} & \multicolumn{3}{c|}{SY}\\
$\alpha$ & $\beta$ & MASE & MMR & U2 & MASE & MMR & U2 & MASE & MMR & U2\\
\hline
0.10 & 0.10 & 0.023 & 0.988 & 0.718 & -0.029 & 0.989 & 0.717 & -0.002 & 0.988 & 0.718\\
0.20 & 0.20 & 0.051 & 0.988 & 0.729 & -0.055 & 0.989 & 0.725 & 0.001 & 0.989 & 0.728\\
0.30 & 0.30 & 0.080 & 0.989 & 0.741 & -0.084 & 0.990 & 0.733 & 0.006 & 0.990 & 0.740\\
\hline
\end{tabular}
\end{center}
\begin{center}
\begin{tabular}{|rr|rrr|rrr|}
\hline
&& \multicolumn{3}{c|}{TSB} & \multicolumn{3}{c|}{HES}\\
$\alpha$ & $\beta$ & MASE & MMR & U2 & MASE & MMR & U2\\
\hline
0.10 & 0.01 & -0.008 & 0.987 & 0.710 & -0.011 & 0.987 & 0.709\\
0.10 & 0.02 & -0.004 & 0.988 & 0.711 & -0.008 & 0.988 & 0.710\\
0.10 & 0.03 & -0.003 & 0.988 & 0.713 & -0.006 & 0.988 & 0.711\\
0.10 & 0.04 & -0.003 & 0.988 & 0.715 & -0.005 & 0.988 & 0.712\\
0.10 & 0.05 & -0.002 & 0.988 & 0.717 & -0.004 & 0.988 & 0.712\\
0.10 & 0.10 & -0.002 & 0.988 & 0.726 & -0.002 & 0.988 & 0.717\\
0.10 & 0.20 & -0.002 & 0.989 & 0.746 & 0.001 & 0.988 & 0.726\\
0.10 & 0.30 & -0.001 & 0.989 & 0.768 & 0.004 & 0.989 & 0.735\\
\hline
0.20 & 0.01 & -0.008 & 0.987 & 0.710 & -0.011 & 0.987 & 0.709\\
0.20 & 0.02 & -0.004 & 0.988 & 0.711 & -0.008 & 0.988 & 0.710\\
0.20 & 0.03 & -0.003 & 0.988 & 0.713 & -0.006 & 0.988 & 0.711\\
0.20 & 0.04 & -0.003 & 0.988 & 0.715 & -0.005 & 0.988 & 0.712\\
0.20 & 0.05 & -0.002 & 0.988 & 0.717 & -0.004 & 0.988 & 0.712\\
0.20 & 0.10 & -0.002 & 0.988 & 0.726 & -0.002 & 0.988 & 0.717\\
0.20 & 0.20 & -0.002 & 0.989 & 0.746 & 0.001 & 0.988 & 0.726\\
0.20 & 0.30 & -0.001 & 0.989 & 0.768 & 0.004 & 0.989 & 0.735\\
\hline
0.30 & 0.01 & -0.008 & 0.987 & 0.710 & -0.011 & 0.987 & 0.709\\
0.30 & 0.02 & -0.004 & 0.988 & 0.711 & -0.008 & 0.987 & 0.710\\
0.30 & 0.03 & -0.003 & 0.988 & 0.713 & -0.006 & 0.988 & 0.711\\
0.30 & 0.04 & -0.003 & 0.988 & 0.715 & -0.005 & 0.988 & 0.712\\
0.30 & 0.05 & -0.002 & 0.988 & 0.717 & -0.004 & 0.988 & 0.712\\
0.30 & 0.10 & -0.002 & 0.988 & 0.726 & -0.002 & 0.988 & 0.717\\
0.30 & 0.20 & -0.002 & 0.989 & 0.746 & 0.001 & 0.988 & 0.726\\
0.30 & 0.30 & -0.001 & 0.989 & 0.768 & 0.004 & 0.989 & 0.735\\
\hline
\end{tabular}
\end{center}
\caption{Logarithmic demand sizes with $\ell=0.001,p_0=0.5$}
\label{sta3}
\end{table}

\begin{table}
\begin{center}
\begin{tabular}{|rr|rrr|rrr|rrr|}
\hline
&& \multicolumn{3}{c|}{CR} & \multicolumn{3}{c|}{SBA} & \multicolumn{3}{c|}{SY}\\
$\alpha$ & $\beta$ & MASE & MMR & U2 & MASE & MMR & U2 & MASE & MMR & U2\\
\hline
0.10 & 0.10 & 0.026 & 1.635 & 0.711 & -0.005 & 1.603 & 0.710 & 0.001 & 1.610 & 0.710\\
0.20 & 0.20 & 0.053 & 1.663 & 0.716 & -0.011 & 1.597 & 0.714 & 0.003 & 1.611 & 0.715\\
0.30 & 0.30 & 0.086 & 1.697 & 0.724 & -0.015 & 1.592 & 0.718 & 0.008 & 1.616 & 0.720\\
\hline
\end{tabular}
\end{center}
\begin{center}
\begin{tabular}{|rr|rrr|rrr|}
\hline
&& \multicolumn{3}{c|}{TSB} & \multicolumn{3}{c|}{HES}\\
$\alpha$ & $\beta$ & MASE & MMR & U2 & MASE & MMR & U2\\
\hline
0.10 & 0.01 & 0.002 & 1.610 & 0.708 & 0.009 & 1.618 & 0.707\\
0.10 & 0.02 & -0.001 & 1.607 & 0.710 & 0.007 & 1.616 & 0.707\\
0.10 & 0.03 & -0.002 & 1.605 & 0.711 & 0.005 & 1.614 & 0.707\\
0.10 & 0.04 & -0.002 & 1.605 & 0.713 & 0.004 & 1.612 & 0.708\\
0.10 & 0.05 & -0.002 & 1.605 & 0.715 & 0.003 & 1.611 & 0.708\\
0.10 & 0.10 & -0.001 & 1.605 & 0.724 & 0.000 & 1.608 & 0.710\\
0.10 & 0.20 & -0.001 & 1.606 & 0.744 & 0.001 & 1.608 & 0.713\\
0.10 & 0.30 & -0.001 & 1.606 & 0.765 & 0.005 & 1.611 & 0.717\\
\hline
0.20 & 0.01 & 0.002 & 1.611 & 0.708 & 0.009 & 1.618 & 0.707\\
0.20 & 0.02 & -0.001 & 1.607 & 0.710 & 0.007 & 1.616 & 0.707\\
0.20 & 0.03 & -0.001 & 1.606 & 0.711 & 0.005 & 1.614 & 0.707\\
0.20 & 0.04 & -0.002 & 1.605 & 0.713 & 0.004 & 1.613 & 0.708\\
0.20 & 0.05 & -0.001 & 1.605 & 0.715 & 0.003 & 1.611 & 0.708\\
0.20 & 0.10 & -0.001 & 1.605 & 0.724 & 0.000 & 1.608 & 0.710\\
0.20 & 0.20 & -0.001 & 1.606 & 0.744 & 0.001 & 1.608 & 0.713\\
0.20 & 0.30 & -0.001 & 1.606 & 0.766 & 0.005 & 1.612 & 0.717\\
\hline
0.30 & 0.01 & 0.002 & 1.611 & 0.708 & 0.010 & 1.618 & 0.707\\
0.30 & 0.02 & -0.000 & 1.607 & 0.710 & 0.007 & 1.616 & 0.707\\
0.30 & 0.03 & -0.001 & 1.606 & 0.711 & 0.006 & 1.614 & 0.708\\
0.30 & 0.04 & -0.001 & 1.605 & 0.713 & 0.004 & 1.613 & 0.708\\
0.30 & 0.05 & -0.001 & 1.605 & 0.715 & 0.003 & 1.612 & 0.708\\
0.30 & 0.10 & -0.000 & 1.606 & 0.724 & 0.001 & 1.609 & 0.710\\
0.30 & 0.20 & -0.001 & 1.606 & 0.744 & 0.001 & 1.608 & 0.713\\
0.30 & 0.30 & -0.001 & 1.606 & 0.766 & 0.005 & 1.612 & 0.717\\
\hline
\end{tabular}
\end{center}
\caption{Logarithmic demand sizes with $\ell=0.001,p_0=0.2$}
\label{sta4}
\end{table}

\begin{table}
\begin{center}
\begin{tabular}{|rr|rr|rr|rr|}
\hline
$\ell$ & $p_0$ & $\alpha_{\mbox{\tiny TSB}}$ & $\beta_{\mbox{\tiny TSB}}$ &
$\alpha_{\mbox{\tiny HES}}$ & $\beta_{\mbox{\tiny HES}}$ & RGRMSE & PB\\
\hline
\multicolumn{8}{|c|}{MMR-best factors}\\
\hline
0.900 & 0.5 & 0.1 & 0.01 & 0.1 & 0.01 & 1.630 & 35\\
0.900 & 0.2 & 0.1 & 0.03 & 0.1 & 0.06 & 2.578 & 20\\
0.001 & 0.5 & 0.1 & 0.01 & 0.1 & 0.01 & 1.001 & 50\\
0.001 & 0.2 & 0.1 & 0.03 & 0.1 & 0.06 & 1.015 & 50\\
\hline
\multicolumn{8}{|c|}{U2-best factors}\\
\hline
0.900 & 0.5 & 0.1 & 0.01 & 0.1 & 0.01 & 0.997 & 53\\
0.900 & 0.2 & 0.1 & 0.01 & 0.1 & 0.01 & 1.011 & 47\\
0.001 & 0.5 & 0.1 & 0.01 & 0.1 & 0.01 & 1.001 & 50\\
0.001 & 0.2 & 0.1 & 0.01 & 0.1 & 0.01 & 1.016 & 47\\
\hline
\end{tabular}
\end{center}
\caption{Comparison of HES and TSB on logarithmic demand sizes}
\label{logcomp}
\end{table}

\begin{table}
\begin{center}
\begin{tabular}{|rr|rrr|rrr|rrr|}
\hline
&& \multicolumn{3}{c|}{CR} & \multicolumn{3}{c|}{SBA} & \multicolumn{3}{c|}{SY}\\
$\alpha$ & $\beta$ & MASE & MMR & U2 & MASE & MMR & U2 & MASE & MMR & U2\\
\hline
0.10 & 0.10 & 0.022 & 1.169 & 0.713 & -0.014 & 1.150 & 0.712 & 0.005 & 1.160 & 0.712\\
0.20 & 0.20 & 0.039 & 1.190 & 0.723 & -0.034 & 1.149 & 0.719 & 0.005 & 1.171 & 0.722\\
0.30 & 0.30 & 0.057 & 1.212 & 0.735 & -0.055 & 1.147 & 0.726 & 0.006 & 1.184 & 0.732\\
\hline
\end{tabular}
\end{center}
\begin{center}
\begin{tabular}{|rr|rrr|rrr|}
\hline
&& \multicolumn{3}{c|}{TSB} & \multicolumn{3}{c|}{HES}\\
$\alpha$ & $\beta$ & MASE & MMR & U2 & MASE & MMR & U2\\
\hline
0.10 & 0.01 & 0.001 & 1.156 & 0.710 & 0.000 & 1.156 & 0.710\\
0.10 & 0.02 & 0.003 & 1.157 & 0.711 & 0.002 & 1.156 & 0.710\\
0.10 & 0.03 & 0.004 & 1.158 & 0.711 & 0.003 & 1.157 & 0.710\\
0.10 & 0.04 & 0.004 & 1.159 & 0.712 & 0.003 & 1.157 & 0.711\\
0.10 & 0.05 & 0.004 & 1.160 & 0.713 & 0.004 & 1.158 & 0.711\\
0.10 & 0.10 & 0.004 & 1.165 & 0.716 & 0.005 & 1.160 & 0.712\\
0.10 & 0.20 & 0.003 & 1.176 & 0.724 & 0.006 & 1.164 & 0.715\\
0.10 & 0.30 & 0.003 & 1.188 & 0.733 & 0.007 & 1.170 & 0.719\\
\hline
0.20 & 0.01 & 0.001 & 1.163 & 0.716 & 0.000 & 1.162 & 0.716\\
0.20 & 0.02 & 0.003 & 1.164 & 0.717 & 0.002 & 1.163 & 0.716\\
0.20 & 0.03 & 0.004 & 1.165 & 0.717 & 0.003 & 1.163 & 0.716\\
0.20 & 0.04 & 0.004 & 1.166 & 0.718 & 0.003 & 1.164 & 0.717\\
0.20 & 0.05 & 0.004 & 1.167 & 0.719 & 0.004 & 1.164 & 0.717\\
0.20 & 0.10 & 0.004 & 1.171 & 0.722 & 0.005 & 1.167 & 0.718\\
0.20 & 0.20 & 0.003 & 1.182 & 0.731 & 0.005 & 1.171 & 0.722\\
0.20 & 0.30 & 0.003 & 1.194 & 0.740 & 0.007 & 1.176 & 0.725\\
\hline
0.30 & 0.01 & 0.001 & 1.170 & 0.723 & 0.000 & 1.170 & 0.722\\
0.30 & 0.02 & 0.003 & 1.171 & 0.723 & 0.001 & 1.170 & 0.723\\
0.30 & 0.03 & 0.003 & 1.172 & 0.724 & 0.002 & 1.171 & 0.723\\
0.30 & 0.04 & 0.004 & 1.173 & 0.725 & 0.003 & 1.172 & 0.723\\
0.30 & 0.05 & 0.004 & 1.174 & 0.725 & 0.004 & 1.172 & 0.724\\
0.30 & 0.10 & 0.003 & 1.179 & 0.729 & 0.004 & 1.174 & 0.725\\
0.30 & 0.20 & 0.003 & 1.189 & 0.738 & 0.005 & 1.178 & 0.728\\
0.30 & 0.30 & 0.002 & 1.201 & 0.747 & 0.007 & 1.184 & 0.732\\
\hline
\end{tabular}
\end{center}
\caption{Geometric demand sizes with $g=0.8,p_0=0.5$}
\label{stu1}
\end{table}

\begin{table}
\begin{center}
\begin{tabular}{|rr|rrr|rrr|rrr|}
\hline
&& \multicolumn{3}{c|}{CR} & \multicolumn{3}{c|}{SBA} & \multicolumn{3}{c|}{SY}\\
$\alpha$ & $\beta$ & MASE & MMR & U2 & MASE & MMR & U2 & MASE & MMR & U2\\
\hline
0.10 & 0.10 & 0.041 & 1.668 & 0.714 & 0.012 & 1.633 & 0.713 & 0.018 & 1.640 & 0.713\\
0.20 & 0.20 & 0.067 & 1.708 & 0.721 & 0.007 & 1.633 & 0.718 & 0.020 & 1.649 & 0.719\\
0.30 & 0.30 & 0.097 & 1.754 & 0.731 & 0.003 & 1.634 & 0.724 & 0.024 & 1.662 & 0.726\\
\hline
\end{tabular}
\end{center}
\begin{center}
\begin{tabular}{|rr|rrr|rrr|}
\hline
&& \multicolumn{3}{c|}{TSB} & \multicolumn{3}{c|}{HES}\\
$\alpha$ & $\beta$ & MASE & MMR & U2 & MASE & MMR & U2\\
\hline
0.10 & 0.01 & 0.020 & 1.641 & 0.712 & 0.023 & 1.645 & 0.712\\
0.10 & 0.02 & 0.017 & 1.639 & 0.713 & 0.024 & 1.645 & 0.712\\
0.10 & 0.03 & 0.016 & 1.639 & 0.714 & 0.023 & 1.644 & 0.712\\
0.10 & 0.04 & 0.015 & 1.639 & 0.715 & 0.022 & 1.643 & 0.712\\
0.10 & 0.05 & 0.014 & 1.639 & 0.716 & 0.021 & 1.642 & 0.712\\
0.10 & 0.10 & 0.012 & 1.641 & 0.720 & 0.018 & 1.640 & 0.713\\
0.10 & 0.20 & 0.010 & 1.648 & 0.731 & 0.018 & 1.642 & 0.715\\
0.10 & 0.30 & 0.009 & 1.655 & 0.743 & 0.020 & 1.648 & 0.717\\
\hline
0.20 & 0.01 & 0.021 & 1.646 & 0.715 & 0.025 & 1.650 & 0.714\\
0.20 & 0.02 & 0.019 & 1.644 & 0.716 & 0.025 & 1.650 & 0.715\\
0.20 & 0.03 & 0.017 & 1.643 & 0.717 & 0.024 & 1.649 & 0.715\\
0.20 & 0.04 & 0.016 & 1.643 & 0.718 & 0.023 & 1.648 & 0.715\\
0.20 & 0.05 & 0.015 & 1.643 & 0.718 & 0.022 & 1.647 & 0.715\\
0.20 & 0.10 & 0.012 & 1.644 & 0.723 & 0.019 & 1.645 & 0.716\\
0.20 & 0.20 & 0.010 & 1.650 & 0.734 & 0.019 & 1.647 & 0.718\\
0.20 & 0.30 & 0.009 & 1.657 & 0.746 & 0.021 & 1.652 & 0.720\\
\hline
0.30 & 0.01 & 0.022 & 1.650 & 0.718 & 0.025 & 1.653 & 0.717\\
0.30 & 0.02 & 0.019 & 1.647 & 0.719 & 0.025 & 1.653 & 0.717\\
0.30 & 0.03 & 0.018 & 1.647 & 0.720 & 0.024 & 1.652 & 0.718\\
0.30 & 0.04 & 0.016 & 1.646 & 0.721 & 0.023 & 1.651 & 0.718\\
0.30 & 0.05 & 0.016 & 1.646 & 0.721 & 0.022 & 1.650 & 0.718\\
0.30 & 0.10 & 0.012 & 1.647 & 0.726 & 0.020 & 1.648 & 0.719\\
0.30 & 0.20 & 0.009 & 1.651 & 0.737 & 0.019 & 1.650 & 0.721\\
0.30 & 0.30 & 0.008 & 1.658 & 0.750 & 0.021 & 1.656 & 0.723\\
\hline
\end{tabular}
\end{center}
\caption{Geometric demand sizes with $g=0.8,p_0=0.2$}
\label{stu2}
\end{table}

\begin{table}
\begin{center}
\begin{tabular}{|rr|rrr|rrr|rrr|}
\hline
&& \multicolumn{3}{c|}{CR} & \multicolumn{3}{c|}{SBA} & \multicolumn{3}{c|}{SY}\\
$\alpha$ & $\beta$ & MASE & MMR & U2 & MASE & MMR & U2 & MASE & MMR & U2\\
\hline
0.10 & 0.10 & 0.021 & 1.014 & 0.716 & -0.022 & 1.013 & 0.715 & -0.000 & 1.013 & 0.716\\
0.20 & 0.20 & 0.043 & 1.018 & 0.727 & -0.045 & 1.014 & 0.724 & 0.001 & 1.017 & 0.726\\
0.30 & 0.30 & 0.067 & 1.029 & 0.739 & -0.070 & 1.016 & 0.731 & 0.003 & 1.024 & 0.737\\
\hline
\end{tabular}
\end{center}
\begin{center}
\begin{tabular}{|rr|rrr|rrr|}
\hline
&& \multicolumn{3}{c|}{TSB} & \multicolumn{3}{c|}{HES}\\
$\alpha$ & $\beta$ & MASE & MMR & U2 & MASE & MMR & U2\\
\hline
0.10 & 0.01 & 0.002 & 1.011 & 0.710 & 0.003 & 1.010 & 0.710\\
0.10 & 0.02 & 0.001 & 1.012 & 0.712 & 0.001 & 1.011 & 0.710\\
0.10 & 0.03 & 0.001 & 1.013 & 0.713 & 0.001 & 1.011 & 0.711\\
0.10 & 0.04 & 0.000 & 1.013 & 0.715 & 0.001 & 1.012 & 0.712\\
0.10 & 0.05 & 0.000 & 1.014 & 0.716 & 0.000 & 1.012 & 0.712\\
0.10 & 0.10 & -0.000 & 1.016 & 0.723 & 0.000 & 1.014 & 0.716\\
0.10 & 0.20 & -0.000 & 1.021 & 0.738 & 0.001 & 1.016 & 0.722\\
0.10 & 0.30 & -0.000 & 1.029 & 0.753 & 0.004 & 1.018 & 0.729\\
\hline
0.20 & 0.01 & 0.002 & 1.011 & 0.713 & 0.003 & 1.010 & 0.713\\
0.20 & 0.02 & 0.001 & 1.012 & 0.715 & 0.001 & 1.011 & 0.713\\
0.20 & 0.03 & 0.001 & 1.013 & 0.716 & 0.001 & 1.011 & 0.714\\
0.20 & 0.04 & 0.000 & 1.014 & 0.718 & 0.001 & 1.012 & 0.715\\
0.20 & 0.05 & 0.000 & 1.014 & 0.719 & 0.000 & 1.012 & 0.715\\
0.20 & 0.10 & -0.000 & 1.017 & 0.726 & -0.000 & 1.014 & 0.719\\
0.20 & 0.20 & -0.000 & 1.024 & 0.741 & 0.001 & 1.017 & 0.725\\
0.20 & 0.30 & -0.000 & 1.034 & 0.757 & 0.004 & 1.020 & 0.732\\
\hline
0.30 & 0.01 & 0.001 & 1.012 & 0.716 & 0.003 & 1.011 & 0.716\\
0.30 & 0.02 & 0.001 & 1.013 & 0.718 & 0.001 & 1.011 & 0.716\\
0.30 & 0.03 & 0.000 & 1.014 & 0.719 & 0.001 & 1.012 & 0.717\\
0.30 & 0.04 & 0.000 & 1.015 & 0.721 & 0.000 & 1.013 & 0.718\\
0.30 & 0.05 & -0.000 & 1.016 & 0.722 & 0.000 & 1.013 & 0.718\\
0.30 & 0.10 & -0.000 & 1.020 & 0.729 & -0.000 & 1.016 & 0.722\\
0.30 & 0.20 & -0.000 & 1.028 & 0.744 & 0.001 & 1.019 & 0.728\\
0.30 & 0.30 & -0.000 & 1.039 & 0.760 & 0.004 & 1.024 & 0.735\\
\hline
\end{tabular}
\end{center}
\caption{Geometric demand sizes with $g=0.2,p_0=0.5$}
\label{stu3}
\end{table}

\begin{table}
\begin{center}
\begin{tabular}{|rr|rrr|rrr|rrr|}
\hline
&& \multicolumn{3}{c|}{CR} & \multicolumn{3}{c|}{SBA} & \multicolumn{3}{c|}{SY}\\
$\alpha$ & $\beta$ & MASE & MMR & U2 & MASE & MMR & U2 & MASE & MMR & U2\\
\hline
0.10 & 0.10 & 0.032 & 1.631 & 0.715 & 0.001 & 1.599 & 0.714 & 0.008 & 1.606 & 0.714\\
0.20 & 0.20 & 0.056 & 1.655 & 0.720 & -0.008 & 1.589 & 0.718 & 0.006 & 1.604 & 0.719\\
0.30 & 0.30 & 0.085 & 1.683 & 0.728 & -0.015 & 1.580 & 0.722 & 0.008 & 1.604 & 0.724\\
\hline
\end{tabular}
\end{center}
\begin{center}
\begin{tabular}{|rr|rrr|rrr|}
\hline
&& \multicolumn{3}{c|}{TSB} & \multicolumn{3}{c|}{HES}\\
$\alpha$ & $\beta$ & MASE & MMR & U2 & MASE & MMR & U2\\
\hline
0.10 & 0.01 & 0.007 & 1.606 & 0.713 & 0.001 & 1.599 & 0.712\\
0.10 & 0.02 & 0.007 & 1.604 & 0.714 & 0.004 & 1.602 & 0.712\\
0.10 & 0.03 & 0.004 & 1.602 & 0.715 & 0.006 & 1.604 & 0.712\\
0.10 & 0.04 & 0.002 & 1.599 & 0.717 & 0.008 & 1.606 & 0.713\\
0.10 & 0.05 & 0.001 & 1.598 & 0.718 & 0.009 & 1.607 & 0.713\\
0.10 & 0.10 & -0.000 & 1.594 & 0.725 & 0.007 & 1.605 & 0.714\\
0.10 & 0.20 & -0.001 & 1.593 & 0.741 & 0.004 & 1.601 & 0.717\\
0.10 & 0.30 & -0.001 & 1.593 & 0.759 & 0.006 & 1.601 & 0.720\\
\hline
0.20 & 0.01 & 0.009 & 1.608 & 0.714 & 0.002 & 1.602 & 0.713\\
0.20 & 0.02 & 0.008 & 1.607 & 0.715 & 0.005 & 1.604 & 0.713\\
0.20 & 0.03 & 0.006 & 1.604 & 0.717 & 0.008 & 1.607 & 0.714\\
0.20 & 0.04 & 0.004 & 1.602 & 0.718 & 0.009 & 1.609 & 0.714\\
0.20 & 0.05 & 0.003 & 1.600 & 0.719 & 0.010 & 1.610 & 0.714\\
0.20 & 0.10 & 0.001 & 1.596 & 0.726 & 0.009 & 1.608 & 0.715\\
0.20 & 0.20 & 0.000 & 1.594 & 0.743 & 0.006 & 1.603 & 0.718\\
0.20 & 0.30 & -0.000 & 1.594 & 0.760 & 0.007 & 1.603 & 0.721\\
\hline
0.30 & 0.01 & 0.010 & 1.609 & 0.715 & 0.003 & 1.602 & 0.714\\
0.30 & 0.02 & 0.009 & 1.608 & 0.716 & 0.005 & 1.605 & 0.714\\
0.30 & 0.03 & 0.006 & 1.605 & 0.718 & 0.008 & 1.607 & 0.715\\
0.30 & 0.04 & 0.005 & 1.602 & 0.719 & 0.010 & 1.609 & 0.715\\
0.30 & 0.05 & 0.003 & 1.600 & 0.720 & 0.011 & 1.610 & 0.715\\
0.30 & 0.10 & 0.001 & 1.596 & 0.728 & 0.010 & 1.609 & 0.717\\
0.30 & 0.20 & 0.000 & 1.594 & 0.744 & 0.007 & 1.604 & 0.719\\
0.30 & 0.30 & 0.000 & 1.595 & 0.762 & 0.008 & 1.604 & 0.722\\
\hline
\end{tabular}
\end{center}
\caption{Geometric demand sizes with $g=0.2,p_0=0.2$}
\label{stu4}
\end{table}

\begin{table}
\begin{center}
\begin{tabular}{|rr|rr|rr|rr|}
\hline
$g$ & $p_0$ & $\alpha_{\mbox{\tiny TSB}}$ & $\beta_{\mbox{\tiny TSB}}$ &
$\alpha_{\mbox{\tiny HES}}$ & $\beta_{\mbox{\tiny HES}}$ & RGRMSE & PB\\
\hline
\multicolumn{8}{|c|}{MMR-best factors}\\
\hline
0.900 & 0.5 & 0.1 & 0.01 & 0.1 & 0.01 & 1.727 & 39\\
0.900 & 0.2 & 0.1 & 0.02 & 0.1 & 0.06 & 3.244 & 20\\
0.001 & 0.5 & 0.1 & 0.01 & 0.1 & 0.01 & 0.983 & 50\\
0.001 & 0.2 & 0.1 & 0.07 & 0.1 & 0.01 & 1.565 & 39\\
\hline
\multicolumn{8}{|c|}{U2-best factors}\\
\hline
0.900 & 0.5 & 0.1 & 0.01 & 0.1 & 0.01 & 1.001 & 51\\
0.900 & 0.2 & 0.1 & 0.01 & 0.1 & 0.01 & 1.004 & 50\\
0.001 & 0.5 & 0.1 & 0.01 & 0.1 & 0.01 & 1.000 & 51\\
0.001 & 0.2 & 0.1 & 0.01 & 0.1 & 0.01 & 1.000 & 53\\
\hline
\end{tabular}
\end{center}
\caption{Comparison of HES and TSB on geometric demand sizes}
\label{geocomp}
\end{table}

\begin{table}
\begin{center}
\begin{tabular}{|rr|rrr|rrr|rrr|}
\hline
&& \multicolumn{3}{c|}{CR} & \multicolumn{3}{c|}{SBA} & \multicolumn{3}{c|}{SY}\\
$\alpha$ & $\beta$ & MASE & MMR & U2 & MASE & MMR & U2 & MASE & MMR & U2\\
\hline
0.10 & 0.10 & 0.161 & 1.785 & 0.722 & 0.136 & 1.736 & 0.719 & 0.146 & 1.754 & 0.720\\
0.20 & 0.20 & 0.135 & 1.735 & 0.728 & 0.087 & 1.642 & 0.721 & 0.105 & 1.676 & 0.724\\
0.30 & 0.30 & 0.131 & 1.737 & 0.738 & 0.060 & 1.599 & 0.725 & 0.088 & 1.651 & 0.731\\
\hline
\end{tabular}
\end{center}
\begin{center}
\begin{tabular}{|rr|rrr|rrr|}
\hline
&& \multicolumn{3}{c|}{TSB} & \multicolumn{3}{c|}{HES}\\
$\alpha$ & $\beta$ & MASE & MMR & U2 & MASE & MMR & U2\\
\hline
0.10 & 0.01 & 0.234 & 1.958 & 0.734 & 0.285 & 2.088 & 0.747\\
0.10 & 0.02 & 0.174 & 1.812 & 0.723 & 0.250 & 2.000 & 0.738\\
0.10 & 0.03 & 0.136 & 1.723 & 0.718 & 0.223 & 1.935 & 0.732\\
0.10 & 0.04 & 0.110 & 1.666 & 0.715 & 0.202 & 1.885 & 0.728\\
0.10 & 0.05 & 0.092 & 1.627 & 0.714 & 0.186 & 1.844 & 0.725\\
0.10 & 0.10 & 0.050 & 1.540 & 0.714 & 0.134 & 1.725 & 0.718\\
0.10 & 0.20 & 0.024 & 1.502 & 0.720 & 0.090 & 1.630 & 0.715\\
0.10 & 0.30 & 0.015 & 1.499 & 0.727 & 0.070 & 1.591 & 0.715\\
\hline
0.20 & 0.01 & 0.235 & 1.973 & 0.744 & 0.287 & 2.105 & 0.759\\
0.20 & 0.02 & 0.175 & 1.825 & 0.731 & 0.252 & 2.016 & 0.749\\
0.20 & 0.03 & 0.137 & 1.735 & 0.725 & 0.225 & 1.950 & 0.742\\
0.20 & 0.04 & 0.111 & 1.677 & 0.722 & 0.204 & 1.899 & 0.737\\
0.20 & 0.05 & 0.094 & 1.638 & 0.720 & 0.187 & 1.858 & 0.733\\
0.20 & 0.10 & 0.051 & 1.550 & 0.720 & 0.135 & 1.738 & 0.725\\
0.20 & 0.20 & 0.025 & 1.511 & 0.726 & 0.092 & 1.642 & 0.721\\
0.20 & 0.30 & 0.016 & 1.508 & 0.734 & 0.072 & 1.602 & 0.721\\
\hline
0.30 & 0.01 & 0.236 & 1.986 & 0.754 & 0.289 & 2.119 & 0.771\\
0.30 & 0.02 & 0.176 & 1.837 & 0.739 & 0.253 & 2.030 & 0.760\\
0.30 & 0.03 & 0.137 & 1.746 & 0.732 & 0.226 & 1.963 & 0.752\\
0.30 & 0.04 & 0.112 & 1.688 & 0.729 & 0.205 & 1.912 & 0.746\\
0.30 & 0.05 & 0.094 & 1.648 & 0.727 & 0.188 & 1.871 & 0.742\\
0.30 & 0.10 & 0.052 & 1.560 & 0.726 & 0.136 & 1.750 & 0.732\\
0.30 & 0.20 & 0.026 & 1.520 & 0.732 & 0.092 & 1.654 & 0.728\\
0.30 & 0.30 & 0.017 & 1.516 & 0.741 & 0.073 & 1.614 & 0.728\\
\hline
\end{tabular}
\end{center}
\caption{Decreasing demand with $\ell=0.9,p_0=0.5$}
\label{dec1}
\end{table}

\begin{table}
\begin{center}
\begin{tabular}{|rr|rrr|rrr|rrr|}
\hline
&& \multicolumn{3}{c|}{CR} & \multicolumn{3}{c|}{SBA} & \multicolumn{3}{c|}{SY}\\
$\alpha$ & $\beta$ & MASE & MMR & U2 & MASE & MMR & U2 & MASE & MMR & U2\\
\hline
0.10 & 0.10 & 0.208 & 2.460 & 0.713 & 0.185 & 2.386 & 0.711 & 0.189 & 2.399 & 0.711\\
0.20 & 0.20 & 0.185 & 2.378 & 0.714 & 0.141 & 2.237 & 0.710 & 0.149 & 2.260 & 0.711\\
0.30 & 0.30 & 0.178 & 2.354 & 0.718 & 0.114 & 2.146 & 0.711 & 0.126 & 2.181 & 0.712\\
\hline
\end{tabular}
\end{center}
\begin{center}
\begin{tabular}{|rr|rrr|rrr|}
\hline
&& \multicolumn{3}{c|}{TSB} & \multicolumn{3}{c|}{HES}\\
$\alpha$ & $\beta$ & MASE & MMR & U2 & MASE & MMR & U2\\
\hline
0.10 & 0.01 & 0.192 & 2.409 & 0.711 & 0.262 & 2.666 & 0.721\\
0.10 & 0.02 & 0.142 & 2.226 & 0.707 & 0.247 & 2.609 & 0.718\\
0.10 & 0.03 & 0.111 & 2.114 & 0.705 & 0.233 & 2.560 & 0.716\\
0.10 & 0.04 & 0.090 & 2.041 & 0.705 & 0.221 & 2.516 & 0.715\\
0.10 & 0.05 & 0.075 & 1.991 & 0.705 & 0.211 & 2.478 & 0.713\\
0.10 & 0.10 & 0.042 & 1.878 & 0.708 & 0.172 & 2.338 & 0.709\\
0.10 & 0.20 & 0.022 & 1.820 & 0.716 & 0.129 & 2.183 & 0.706\\
0.10 & 0.30 & 0.015 & 1.805 & 0.725 & 0.105 & 2.101 & 0.706\\
\hline
0.20 & 0.01 & 0.193 & 2.417 & 0.715 & 0.263 & 2.676 & 0.726\\
0.20 & 0.02 & 0.142 & 2.230 & 0.710 & 0.247 & 2.619 & 0.723\\
0.20 & 0.03 & 0.110 & 2.116 & 0.708 & 0.234 & 2.569 & 0.721\\
0.20 & 0.04 & 0.089 & 2.041 & 0.707 & 0.222 & 2.525 & 0.719\\
0.20 & 0.05 & 0.074 & 1.991 & 0.707 & 0.211 & 2.486 & 0.717\\
0.20 & 0.10 & 0.040 & 1.878 & 0.710 & 0.172 & 2.342 & 0.712\\
0.20 & 0.20 & 0.021 & 1.821 & 0.719 & 0.128 & 2.184 & 0.709\\
0.20 & 0.30 & 0.014 & 1.807 & 0.729 & 0.104 & 2.100 & 0.708\\
\hline
0.30 & 0.01 & 0.193 & 2.424 & 0.719 & 0.264 & 2.687 & 0.732\\
0.30 & 0.02 & 0.141 & 2.234 & 0.713 & 0.248 & 2.629 & 0.729\\
0.30 & 0.03 & 0.109 & 2.118 & 0.711 & 0.235 & 2.578 & 0.726\\
0.30 & 0.04 & 0.088 & 2.043 & 0.710 & 0.223 & 2.533 & 0.724\\
0.30 & 0.05 & 0.073 & 1.991 & 0.710 & 0.212 & 2.494 & 0.722\\
0.30 & 0.10 & 0.039 & 1.878 & 0.713 & 0.172 & 2.347 & 0.716\\
0.30 & 0.20 & 0.020 & 1.823 & 0.722 & 0.127 & 2.186 & 0.711\\
0.30 & 0.30 & 0.014 & 1.809 & 0.733 & 0.103 & 2.099 & 0.710\\
\hline
\end{tabular}
\end{center}
\caption{Decreasing demand with $\ell=0.9,p_0=0.2$}
\label{dec2}
\end{table}

\begin{table}
\begin{center}
\begin{tabular}{|rr|rrr|rrr|rrr|}
\hline
&& \multicolumn{3}{c|}{CR} & \multicolumn{3}{c|}{SBA} & \multicolumn{3}{c|}{SY}\\
$\alpha$ & $\beta$ & MASE & MMR & U2 & MASE & MMR & U2 & MASE & MMR & U2\\
\hline
0.10 & 0.10 & 0.243 & 1.638 & 0.748 & 0.206 & 1.606 & 0.740 & 0.221 & 1.616 & 0.743\\
0.20 & 0.20 & 0.201 & 1.576 & 0.744 & 0.131 & 1.518 & 0.731 & 0.158 & 1.536 & 0.736\\
0.30 & 0.30 & 0.194 & 1.559 & 0.750 & 0.091 & 1.475 & 0.731 & 0.131 & 1.501 & 0.739\\
\hline
\end{tabular}
\end{center}
\begin{center}
\begin{tabular}{|rr|rrr|rrr|}
\hline
&& \multicolumn{3}{c|}{TSB} & \multicolumn{3}{c|}{HES}\\
$\alpha$ & $\beta$ & MASE & MMR & U2 & MASE & MMR & U2\\
\hline
0.10 & 0.01 & 0.352 & 1.791 & 0.784 & 0.428 & 1.905 & 0.822\\
0.10 & 0.02 & 0.263 & 1.661 & 0.750 & 0.376 & 1.829 & 0.795\\
0.10 & 0.03 & 0.207 & 1.583 & 0.736 & 0.337 & 1.772 & 0.778\\
0.10 & 0.04 & 0.169 & 1.532 & 0.729 & 0.305 & 1.728 & 0.766\\
0.10 & 0.05 & 0.142 & 1.497 & 0.726 & 0.280 & 1.694 & 0.757\\
0.10 & 0.10 & 0.079 & 1.421 & 0.728 & 0.203 & 1.590 & 0.737\\
0.10 & 0.20 & 0.042 & 1.381 & 0.746 & 0.138 & 1.507 & 0.729\\
0.10 & 0.30 & 0.028 & 1.368 & 0.767 & 0.109 & 1.471 & 0.731\\
\hline
0.20 & 0.01 & 0.352 & 1.790 & 0.784 & 0.428 & 1.905 & 0.822\\
0.20 & 0.02 & 0.263 & 1.661 & 0.750 & 0.376 & 1.829 & 0.795\\
0.20 & 0.03 & 0.207 & 1.582 & 0.736 & 0.336 & 1.772 & 0.778\\
0.20 & 0.04 & 0.169 & 1.532 & 0.729 & 0.305 & 1.728 & 0.766\\
0.20 & 0.05 & 0.142 & 1.497 & 0.726 & 0.280 & 1.694 & 0.757\\
0.20 & 0.10 & 0.079 & 1.421 & 0.728 & 0.203 & 1.590 & 0.737\\
0.20 & 0.20 & 0.042 & 1.380 & 0.746 & 0.138 & 1.507 & 0.729\\
0.20 & 0.30 & 0.028 & 1.368 & 0.767 & 0.109 & 1.471 & 0.731\\
\hline
0.30 & 0.01 & 0.352 & 1.790 & 0.784 & 0.428 & 1.905 & 0.822\\
0.30 & 0.02 & 0.263 & 1.661 & 0.750 & 0.376 & 1.829 & 0.795\\
0.30 & 0.03 & 0.207 & 1.582 & 0.736 & 0.336 & 1.772 & 0.778\\
0.30 & 0.04 & 0.169 & 1.532 & 0.729 & 0.305 & 1.728 & 0.766\\
0.30 & 0.05 & 0.142 & 1.497 & 0.726 & 0.280 & 1.694 & 0.757\\
0.30 & 0.10 & 0.079 & 1.420 & 0.728 & 0.203 & 1.590 & 0.737\\
0.30 & 0.20 & 0.042 & 1.380 & 0.746 & 0.138 & 1.507 & 0.729\\
0.30 & 0.30 & 0.028 & 1.367 & 0.767 & 0.109 & 1.471 & 0.731\\
\hline
\end{tabular}
\end{center}
\caption{Decreasing demand with $\ell=0.001,p_0=0.5$}
\label{dec3}
\end{table}

\begin{table}
\begin{center}
\begin{tabular}{|rr|rrr|rrr|rrr|}
\hline
&& \multicolumn{3}{c|}{CR} & \multicolumn{3}{c|}{SBA} & \multicolumn{3}{c|}{SY}\\
$\alpha$ & $\beta$ & MASE & MMR & U2 & MASE & MMR & U2 & MASE & MMR & U2\\
\hline
0.10 & 0.10 & 0.234 & 2.439 & 0.738 & 0.208 & 2.367 & 0.732 & 0.213 & 2.380 & 0.733\\
0.20 & 0.20 & 0.210 & 2.358 & 0.735 & 0.161 & 2.222 & 0.726 & 0.169 & 2.245 & 0.728\\
0.30 & 0.30 & 0.207 & 2.339 & 0.740 & 0.133 & 2.138 & 0.725 & 0.146 & 2.172 & 0.728\\
\hline
\end{tabular}
\end{center}
\begin{center}
\begin{tabular}{|rr|rrr|rrr|}
\hline
&& \multicolumn{3}{c|}{TSB} & \multicolumn{3}{c|}{HES}\\
$\alpha$ & $\beta$ & MASE & MMR & U2 & MASE & MMR & U2\\
\hline
0.10 & 0.01 & 0.217 & 2.391 & 0.732 & 0.296 & 2.643 & 0.756\\
0.10 & 0.02 & 0.160 & 2.211 & 0.721 & 0.278 & 2.587 & 0.750\\
0.10 & 0.03 & 0.125 & 2.101 & 0.717 & 0.263 & 2.538 & 0.745\\
0.10 & 0.04 & 0.101 & 2.029 & 0.716 & 0.249 & 2.495 & 0.741\\
0.10 & 0.05 & 0.085 & 1.979 & 0.717 & 0.237 & 2.458 & 0.738\\
0.10 & 0.10 & 0.046 & 1.865 & 0.724 & 0.194 & 2.320 & 0.728\\
0.10 & 0.20 & 0.023 & 1.800 & 0.743 & 0.145 & 2.169 & 0.721\\
0.10 & 0.30 & 0.015 & 1.777 & 0.765 & 0.119 & 2.088 & 0.720\\
\hline
0.20 & 0.01 & 0.217 & 2.391 & 0.732 & 0.296 & 2.643 & 0.756\\
0.20 & 0.02 & 0.160 & 2.212 & 0.721 & 0.278 & 2.587 & 0.750\\
0.20 & 0.03 & 0.125 & 2.102 & 0.717 & 0.263 & 2.538 & 0.745\\
0.20 & 0.04 & 0.101 & 2.029 & 0.716 & 0.249 & 2.495 & 0.741\\
0.20 & 0.05 & 0.085 & 1.979 & 0.717 & 0.237 & 2.458 & 0.738\\
0.20 & 0.10 & 0.046 & 1.865 & 0.724 & 0.194 & 2.320 & 0.728\\
0.20 & 0.20 & 0.023 & 1.800 & 0.743 & 0.145 & 2.169 & 0.721\\
0.20 & 0.30 & 0.015 & 1.777 & 0.765 & 0.119 & 2.088 & 0.720\\
\hline
0.30 & 0.01 & 0.217 & 2.390 & 0.732 & 0.296 & 2.643 & 0.756\\
0.30 & 0.02 & 0.160 & 2.211 & 0.721 & 0.278 & 2.587 & 0.750\\
0.30 & 0.03 & 0.125 & 2.101 & 0.717 & 0.263 & 2.538 & 0.745\\
0.30 & 0.04 & 0.101 & 2.029 & 0.716 & 0.249 & 2.495 & 0.741\\
0.30 & 0.05 & 0.085 & 1.979 & 0.717 & 0.237 & 2.457 & 0.738\\
0.30 & 0.10 & 0.046 & 1.865 & 0.724 & 0.194 & 2.320 & 0.728\\
0.30 & 0.20 & 0.023 & 1.800 & 0.743 & 0.145 & 2.169 & 0.721\\
0.30 & 0.30 & 0.015 & 1.777 & 0.765 & 0.119 & 2.088 & 0.720\\
\hline
\end{tabular}
\end{center}
\caption{Decreasing demand with $\ell=0.001,p_0=0.2$}
\label{dec4}
\end{table}

\begin{table}
\begin{center}
\begin{tabular}{|rr|rrr|rrr|rrr|}
\hline
&& \multicolumn{3}{c|}{CR} & \multicolumn{3}{c|}{SBA} & \multicolumn{3}{c|}{SY}\\
$\alpha$ & $\beta$ & MASE & MMR & U2 & MASE & MMR & U2 & MASE & MMR & U2\\
\hline
0.10 & 0.10 & 0.348 & 2.216 & 0.800 & 0.313 & 2.142 & 0.791 & 0.331 & 2.180 & 0.795\\
0.20 & 0.20 & 0.359 & 2.254 & 0.820 & 0.289 & 2.101 & 0.799 & 0.326 & 2.182 & 0.810\\
0.30 & 0.30 & 0.375 & 2.309 & 0.846 & 0.268 & 2.073 & 0.809 & 0.326 & 2.203 & 0.831\\
\hline
\end{tabular}
\end{center}
\begin{center}
\begin{tabular}{|rr|rrr|rrr|}
\hline
&& \multicolumn{3}{c|}{TSB} & \multicolumn{3}{c|}{HES}\\
$\alpha$ & $\beta$ & MASE & MMR & U2 & MASE & MMR & U2\\
\hline
0.10 & 0.01 & 0.249 & 1.937 & 0.760 & 0.306 & 2.106 & 0.781\\
0.10 & 0.02 & 0.195 & 1.779 & 0.744 & 0.289 & 2.056 & 0.775\\
0.10 & 0.03 & 0.156 & 1.667 & 0.736 & 0.275 & 2.013 & 0.769\\
0.10 & 0.04 & 0.128 & 1.587 & 0.731 & 0.262 & 1.975 & 0.764\\
0.10 & 0.05 & 0.108 & 1.528 & 0.728 & 0.250 & 1.942 & 0.761\\
0.10 & 0.10 & 0.059 & 1.389 & 0.724 & 0.208 & 1.821 & 0.748\\
0.10 & 0.20 & 0.032 & 1.322 & 0.727 & 0.164 & 1.691 & 0.740\\
0.10 & 0.30 & 0.023 & 1.309 & 0.732 & 0.141 & 1.626 & 0.738\\
\hline
0.20 & 0.01 & 0.243 & 1.935 & 0.769 & 0.300 & 2.101 & 0.792\\
0.20 & 0.02 & 0.190 & 1.778 & 0.753 & 0.283 & 2.052 & 0.785\\
0.20 & 0.03 & 0.152 & 1.668 & 0.744 & 0.269 & 2.009 & 0.779\\
0.20 & 0.04 & 0.125 & 1.589 & 0.739 & 0.256 & 1.972 & 0.774\\
0.20 & 0.05 & 0.105 & 1.532 & 0.736 & 0.244 & 1.939 & 0.770\\
0.20 & 0.10 & 0.057 & 1.396 & 0.732 & 0.203 & 1.819 & 0.757\\
0.20 & 0.20 & 0.030 & 1.331 & 0.735 & 0.160 & 1.694 & 0.748\\
0.20 & 0.30 & 0.021 & 1.318 & 0.741 & 0.137 & 1.631 & 0.747\\
\hline
0.30 & 0.01 & 0.240 & 1.938 & 0.779 & 0.296 & 2.104 & 0.803\\
0.30 & 0.02 & 0.187 & 1.783 & 0.762 & 0.279 & 2.054 & 0.795\\
0.30 & 0.03 & 0.149 & 1.674 & 0.753 & 0.265 & 2.012 & 0.789\\
0.30 & 0.04 & 0.122 & 1.596 & 0.748 & 0.252 & 1.975 & 0.784\\
0.30 & 0.05 & 0.103 & 1.540 & 0.745 & 0.241 & 1.942 & 0.780\\
0.30 & 0.10 & 0.056 & 1.407 & 0.741 & 0.200 & 1.824 & 0.767\\
0.30 & 0.20 & 0.030 & 1.343 & 0.745 & 0.157 & 1.701 & 0.758\\
0.30 & 0.30 & 0.021 & 1.330 & 0.751 & 0.135 & 1.639 & 0.757\\
\hline
\end{tabular}
\end{center}
\caption{Sudden obsolescence with $\ell=0.9,p_0=0.5$}
\label{obs1}
\end{table}

\begin{table}
\begin{center}
\begin{tabular}{|rr|rrr|rrr|rrr|}
\hline
&& \multicolumn{3}{c|}{CR} & \multicolumn{3}{c|}{SBA} & \multicolumn{3}{c|}{SY}\\
$\alpha$ & $\beta$ & MASE & MMR & U2 & MASE & MMR & U2 & MASE & MMR & U2\\
\hline
0.10 & 0.10 & 0.285 & 2.718 & 0.732 & 0.258 & 2.631 & 0.729 & 0.264 & 2.649 & 0.729\\
0.20 & 0.20 & 0.300 & 2.773 & 0.743 & 0.245 & 2.591 & 0.734 & 0.257 & 2.630 & 0.736\\
0.30 & 0.30 & 0.318 & 2.838 & 0.757 & 0.233 & 2.554 & 0.740 & 0.252 & 2.619 & 0.745\\
\hline
\end{tabular}
\end{center}
\begin{center}
\begin{tabular}{|rr|rrr|rrr|}
\hline
&& \multicolumn{3}{c|}{TSB} & \multicolumn{3}{c|}{HES}\\
$\alpha$ & $\beta$ & MASE & MMR & U2 & MASE & MMR & U2\\
\hline
0.10 & 0.01 & 0.202 & 2.403 & 0.716 & 0.264 & 2.647 & 0.727\\
0.10 & 0.02 & 0.155 & 2.220 & 0.710 & 0.256 & 2.615 & 0.725\\
0.10 & 0.03 & 0.124 & 2.096 & 0.708 & 0.248 & 2.586 & 0.724\\
0.10 & 0.04 & 0.101 & 2.008 & 0.707 & 0.241 & 2.559 & 0.723\\
0.10 & 0.05 & 0.085 & 1.944 & 0.706 & 0.235 & 2.534 & 0.722\\
0.10 & 0.10 & 0.046 & 1.799 & 0.708 & 0.210 & 2.434 & 0.718\\
0.10 & 0.20 & 0.025 & 1.730 & 0.714 & 0.177 & 2.302 & 0.714\\
0.10 & 0.30 & 0.018 & 1.716 & 0.722 & 0.156 & 2.221 & 0.713\\
\hline
0.20 & 0.01 & 0.200 & 2.401 & 0.719 & 0.261 & 2.643 & 0.730\\
0.20 & 0.02 & 0.153 & 2.219 & 0.714 & 0.253 & 2.611 & 0.729\\
0.20 & 0.03 & 0.122 & 2.095 & 0.711 & 0.246 & 2.582 & 0.728\\
0.20 & 0.04 & 0.099 & 2.007 & 0.710 & 0.239 & 2.555 & 0.726\\
0.20 & 0.05 & 0.083 & 1.944 & 0.709 & 0.232 & 2.531 & 0.725\\
0.20 & 0.10 & 0.045 & 1.802 & 0.711 & 0.207 & 2.430 & 0.721\\
0.20 & 0.20 & 0.024 & 1.735 & 0.718 & 0.174 & 2.297 & 0.717\\
0.20 & 0.30 & 0.017 & 1.721 & 0.727 & 0.153 & 2.215 & 0.716\\
\hline
0.30 & 0.01 & 0.197 & 2.398 & 0.724 & 0.258 & 2.637 & 0.735\\
0.30 & 0.02 & 0.151 & 2.218 & 0.717 & 0.250 & 2.606 & 0.734\\
0.30 & 0.03 & 0.120 & 2.095 & 0.715 & 0.243 & 2.577 & 0.732\\
0.30 & 0.04 & 0.098 & 2.008 & 0.713 & 0.236 & 2.551 & 0.731\\
0.30 & 0.05 & 0.082 & 1.945 & 0.713 & 0.230 & 2.526 & 0.730\\
0.30 & 0.10 & 0.044 & 1.804 & 0.715 & 0.204 & 2.426 & 0.726\\
0.30 & 0.20 & 0.024 & 1.740 & 0.723 & 0.171 & 2.293 & 0.721\\
0.30 & 0.30 & 0.017 & 1.727 & 0.733 & 0.150 & 2.211 & 0.720\\
\hline
\end{tabular}
\end{center}
\caption{Sudden obsolescence with $\ell=0.9,p_0=0.2$}
\label{obs2}
\end{table}

\begin{table}
\begin{center}
\begin{tabular}{|rr|rrr|rrr|rrr|}
\hline
&& \multicolumn{3}{c|}{CR} & \multicolumn{3}{c|}{SBA} & \multicolumn{3}{c|}{SY}\\
$\alpha$ & $\beta$ & MASE & MMR & U2 & MASE & MMR & U2 & MASE & MMR & U2\\
\hline
0.10 & 0.10 & 0.527 & 2.018 & 1.012 & 0.476 & 1.968 & 0.986 & 0.502 & 1.994 & 1.000\\
0.20 & 0.20 & 0.556 & 2.050 & 1.042 & 0.450 & 1.945 & 0.986 & 0.506 & 2.001 & 1.017\\
0.30 & 0.30 & 0.588 & 2.087 & 1.075 & 0.424 & 1.924 & 0.986 & 0.514 & 2.015 & 1.040\\
\hline
\end{tabular}
\end{center}
\begin{center}
\begin{tabular}{|rr|rrr|rrr|}
\hline
&& \multicolumn{3}{c|}{TSB} & \multicolumn{3}{c|}{HES}\\
$\alpha$ & $\beta$ & MASE & MMR & U2 & MASE & MMR & U2\\
\hline
0.10 & 0.01 & 0.376 & 1.747 & 0.881 & 0.464 & 1.923 & 0.955\\
0.10 & 0.02 & 0.293 & 1.580 & 0.825 & 0.438 & 1.870 & 0.932\\
0.10 & 0.03 & 0.234 & 1.461 & 0.794 & 0.416 & 1.824 & 0.913\\
0.10 & 0.04 & 0.192 & 1.377 & 0.776 & 0.396 & 1.784 & 0.897\\
0.10 & 0.05 & 0.160 & 1.314 & 0.765 & 0.378 & 1.748 & 0.884\\
0.10 & 0.10 & 0.085 & 1.166 & 0.747 & 0.314 & 1.617 & 0.840\\
0.10 & 0.20 & 0.043 & 1.087 & 0.755 & 0.245 & 1.474 & 0.807\\
0.10 & 0.30 & 0.029 & 1.062 & 0.772 & 0.209 & 1.396 & 0.797\\
\hline
0.20 & 0.01 & 0.376 & 1.747 & 0.881 & 0.464 & 1.923 & 0.955\\
0.20 & 0.02 & 0.293 & 1.580 & 0.825 & 0.438 & 1.870 & 0.932\\
0.20 & 0.03 & 0.234 & 1.461 & 0.794 & 0.416 & 1.824 & 0.913\\
0.20 & 0.04 & 0.192 & 1.376 & 0.776 & 0.396 & 1.784 & 0.897\\
0.20 & 0.05 & 0.160 & 1.314 & 0.765 & 0.378 & 1.748 & 0.884\\
0.20 & 0.10 & 0.085 & 1.166 & 0.747 & 0.314 & 1.617 & 0.840\\
0.20 & 0.20 & 0.043 & 1.087 & 0.755 & 0.245 & 1.474 & 0.807\\
0.20 & 0.30 & 0.029 & 1.062 & 0.772 & 0.209 & 1.396 & 0.797\\
\hline
0.30 & 0.01 & 0.376 & 1.746 & 0.881 & 0.464 & 1.923 & 0.955\\
0.30 & 0.02 & 0.293 & 1.579 & 0.825 & 0.438 & 1.870 & 0.932\\
0.30 & 0.03 & 0.234 & 1.461 & 0.794 & 0.416 & 1.824 & 0.913\\
0.30 & 0.04 & 0.192 & 1.376 & 0.776 & 0.396 & 1.784 & 0.897\\
0.30 & 0.05 & 0.160 & 1.314 & 0.765 & 0.378 & 1.748 & 0.884\\
0.30 & 0.10 & 0.085 & 1.166 & 0.747 & 0.314 & 1.617 & 0.840\\
0.30 & 0.20 & 0.043 & 1.087 & 0.755 & 0.245 & 1.474 & 0.807\\
0.30 & 0.30 & 0.029 & 1.062 & 0.772 & 0.209 & 1.396 & 0.797\\
\hline
\end{tabular}
\end{center}
\caption{Sudden obsolescence with $\ell=0.001,p_0=0.5$}
\label{obs3}
\end{table}

\begin{table}
\begin{center}
\begin{tabular}{|rr|rrr|rrr|rrr|}
\hline
&& \multicolumn{3}{c|}{CR} & \multicolumn{3}{c|}{SBA} & \multicolumn{3}{c|}{SY}\\
$\alpha$ & $\beta$ & MASE & MMR & U2 & MASE & MMR & U2 & MASE & MMR & U2\\
\hline
0.10 & 0.10 & 0.325 & 2.705 & 0.797 & 0.294 & 2.619 & 0.788 & 0.301 & 2.637 & 0.790\\
0.20 & 0.20 & 0.349 & 2.768 & 0.814 & 0.285 & 2.591 & 0.793 & 0.299 & 2.630 & 0.798\\
0.30 & 0.30 & 0.377 & 2.842 & 0.834 & 0.277 & 2.566 & 0.798 & 0.300 & 2.629 & 0.808\\
\hline
\end{tabular}
\end{center}
\begin{center}
\begin{tabular}{|rr|rrr|rrr|}
\hline
&& \multicolumn{3}{c|}{TSB} & \multicolumn{3}{c|}{HES}\\
$\alpha$ & $\beta$ & MASE & MMR & U2 & MASE & MMR & U2\\
\hline
0.10 & 0.01 & 0.229 & 2.391 & 0.753 & 0.301 & 2.636 & 0.783\\
0.10 & 0.02 & 0.176 & 2.208 & 0.737 & 0.291 & 2.604 & 0.778\\
0.10 & 0.03 & 0.139 & 2.082 & 0.730 & 0.282 & 2.574 & 0.775\\
0.10 & 0.04 & 0.113 & 1.993 & 0.726 & 0.274 & 2.547 & 0.772\\
0.10 & 0.05 & 0.094 & 1.928 & 0.725 & 0.267 & 2.522 & 0.769\\
0.10 & 0.10 & 0.049 & 1.774 & 0.727 & 0.238 & 2.422 & 0.758\\
0.10 & 0.20 & 0.024 & 1.692 & 0.744 & 0.201 & 2.288 & 0.748\\
0.10 & 0.30 & 0.016 & 1.666 & 0.765 & 0.177 & 2.203 & 0.744\\
\hline
0.20 & 0.01 & 0.229 & 2.391 & 0.753 & 0.301 & 2.636 & 0.782\\
0.20 & 0.02 & 0.176 & 2.208 & 0.737 & 0.291 & 2.604 & 0.778\\
0.20 & 0.03 & 0.139 & 2.082 & 0.730 & 0.282 & 2.574 & 0.775\\
0.20 & 0.04 & 0.113 & 1.993 & 0.726 & 0.274 & 2.547 & 0.771\\
0.20 & 0.05 & 0.095 & 1.928 & 0.725 & 0.267 & 2.522 & 0.769\\
0.20 & 0.10 & 0.049 & 1.774 & 0.727 & 0.238 & 2.422 & 0.758\\
0.20 & 0.20 & 0.024 & 1.692 & 0.744 & 0.201 & 2.288 & 0.748\\
0.20 & 0.30 & 0.016 & 1.666 & 0.765 & 0.177 & 2.203 & 0.744\\
\hline
0.30 & 0.01 & 0.229 & 2.391 & 0.753 & 0.301 & 2.636 & 0.782\\
0.30 & 0.02 & 0.176 & 2.208 & 0.737 & 0.291 & 2.604 & 0.778\\
0.30 & 0.03 & 0.139 & 2.082 & 0.730 & 0.282 & 2.574 & 0.775\\
0.30 & 0.04 & 0.113 & 1.993 & 0.726 & 0.274 & 2.547 & 0.771\\
0.30 & 0.05 & 0.094 & 1.928 & 0.725 & 0.267 & 2.522 & 0.769\\
0.30 & 0.10 & 0.049 & 1.774 & 0.727 & 0.238 & 2.422 & 0.758\\
0.30 & 0.20 & 0.024 & 1.692 & 0.744 & 0.201 & 2.288 & 0.748\\
0.30 & 0.30 & 0.016 & 1.666 & 0.765 & 0.177 & 2.203 & 0.744\\
\hline
\end{tabular}
\end{center}
\caption{Sudden obsolescence with $\ell=0.001,p_0=0.2$}
\label{obs4}
\end{table}

\end{document}